%% file: main.tex
\documentclass[11pt]{article}

\usepackage[numbers,compress]{natbib}

\usepackage[pdftex]{hyperref}
\hypersetup{
    unicode=false,          
    pdftoolbar=true,        
    pdfmenubar=true,        
    pdffitwindow=false,      
    pdfnewwindow=true,      
    colorlinks=true,       
    linkcolor=DarkBlue,          
    citecolor=DarkGreen,        
    filecolor=DarkRed,      
    urlcolor=DarkBlue,          
    pdftitle={},
    pdfauthor={},
}
\usepackage[skip=2mm,indent=5mm]{parskip}
\usepackage[small]{caption}
\RequirePackage{amsthm,amsmath,amsfonts,amssymb}

\RequirePackage{graphicx}
\usepackage{subcaption}
\usepackage{multicol}
\usepackage[svgnames]{xcolor}

\usepackage{algorithm}
\usepackage{algpseudocode}

\usepackage{fullpage}
\usepackage[utf8]{inputenc}
\usepackage[T1]{fontenc}
\usepackage{microtype}
\usepackage{csquotes}
\usepackage[osf,sc]{mathpazo}
\RequirePackage[scaled=0.90]{helvet}
\RequirePackage[scaled=0.85]{beramono}
\RequirePackage{textcomp}

\theoremstyle{plain}
\newtheorem{theorem}{Theorem}

\theoremstyle{definition}
\newtheorem{definition}[theorem]{Definition}

\input{macros}

\usepackage{tikz}
\usepackage{subcaption}
\usepackage{enumitem}

\usepackage{booktabs}

\title{Algorithmic Collective Action in Recommender Systems:\\Promoting Songs by Reordering Playlists}

\makeatletter
\newcommand{\printfnsymbol}[1]{%
  \textsuperscript{\@fnsymbol{#1}}%
}
\makeatother

\usepackage{authblk}
\author[1]{Joachim Baumann\thanks{Work completed while at the Max-Planck Institute for Intelligent Systems, Tübingen.}}
\author[2,3]{Celestine Mendler-Dünner}
\affil[1]{University of Zurich}
\affil[2]{ELLIS Institute, Tübingen}
\affil[3]{Max-Planck Institute for Intelligent Systems, Tübingen and Tübingen AI Center}
\date{}

\begin{document}

\maketitle

\begin{abstract}
We investigate algorithmic collective action in transformer-based recommender systems. Our use case is a music streaming platform where a collective of fans aims to promote the visibility of an underrepresented artist by strategically placing one of their songs in the existing playlists they control. We introduce two easily implementable strategies to select the position at which to insert the song with the goal to boost recommendations at test time. The strategies exploit statistical properties of the learner by targeting discontinuities in the recommendations, and leveraging the long-tail nature of song distributions. We evaluate the efficacy of our strategies using a publicly available recommender system model released by a major music streaming platform. Our findings reveal that through strategic placement even small collectives (controlling less than 0.01\% of the training data) can achieve up to $40\times$ more test time recommendations than an average song with the same number of training set occurrences. Focusing on the externalities of the strategy, we find that the recommendations of other songs are largely preserved, and the newly gained recommendations are distributed across various artists. Together, our findings demonstrate how carefully designed collective action strategies can be effective while not necessarily being adversarial.
\end{abstract}

\section{Introduction}

In the ever-evolving landscape of music discovery, the challenge of accessing and sifting through the overwhelming number of tracks released daily has become increasingly difficult.
This has resulted in a strong dependence on platforms like Spotify, Deezer, and Apple Music, which distribute and promote music through algorithmic song recommendations. These systems rely on historical data to learn user preferences and predict future content consumption~\citep{Hansen2020SpotifyContextualRecommendation,Tomasi2023SpotifyAPCwithRL,Moor2023ExploitingSequential,Bendada2023DeezerPlaylistContinuationTransformers,Bendada2023TrackMix}.

It has been widely documented that music recommendation systems suffer from popularity bias as they tend to concentrate recommendation exposure on a limited fraction of artists, often overlooking new and emerging talent~\citep{Napoli2016LongTail,bauer2017_nonsuperstars_bled,bauer2019_equalopportunities,Coelho2019Digitalmusic,Blake2020data,Jannach2023WhatRecommenders}.
As the success and visibility of artists are deeply influenced by the algorithms of these platforms, this can lead to a considerable imbalance in the music industry~\citep{Aguiar2021Platforms,Prey2022Platformpop} and reinforce existing inequalities~\citep{Tofalvy2023SplendidIsolation}.
Thus, artists have started to fight for more transparency and fairer payments from online streaming services. The ``Justice at Spotify'' campaign, launched by the~\citet{justiceatspotify}, has been signed by more than 28,000 artists. 
At the same time, the International Society for Music Information Retrieval has been arguing for promoting the discovery of less popular artists by recommending `long-tail' items~\citep{bauer2020_helpartists}, as have other researchers~\citep{celma2009music,turnbull_fivetag_ISMIR08,Craw2015MusicRecommendation,Porcaro2024}.

In this work, we explore algorithmic collective action as an alternative lever for emerging artists to gain exposure in machine learning-powered recommender systems by mobilizing their fan base. Algorithmic collective action~\citep{hardt2023algorithmic} refers to the coordinated effort of a group of platform participants who strategically report the part of the training data they control to achieve more favorable prediction outcomes. Our work is situated in an emerging literature that recognizes data as a lever for users to promote their interests on digital platforms~\citep{Vincent2021DataLeverage, hardt2023algorithmic}.

\begin{figure}[t!]
\centering
\includegraphics[width=0.6\textwidth]{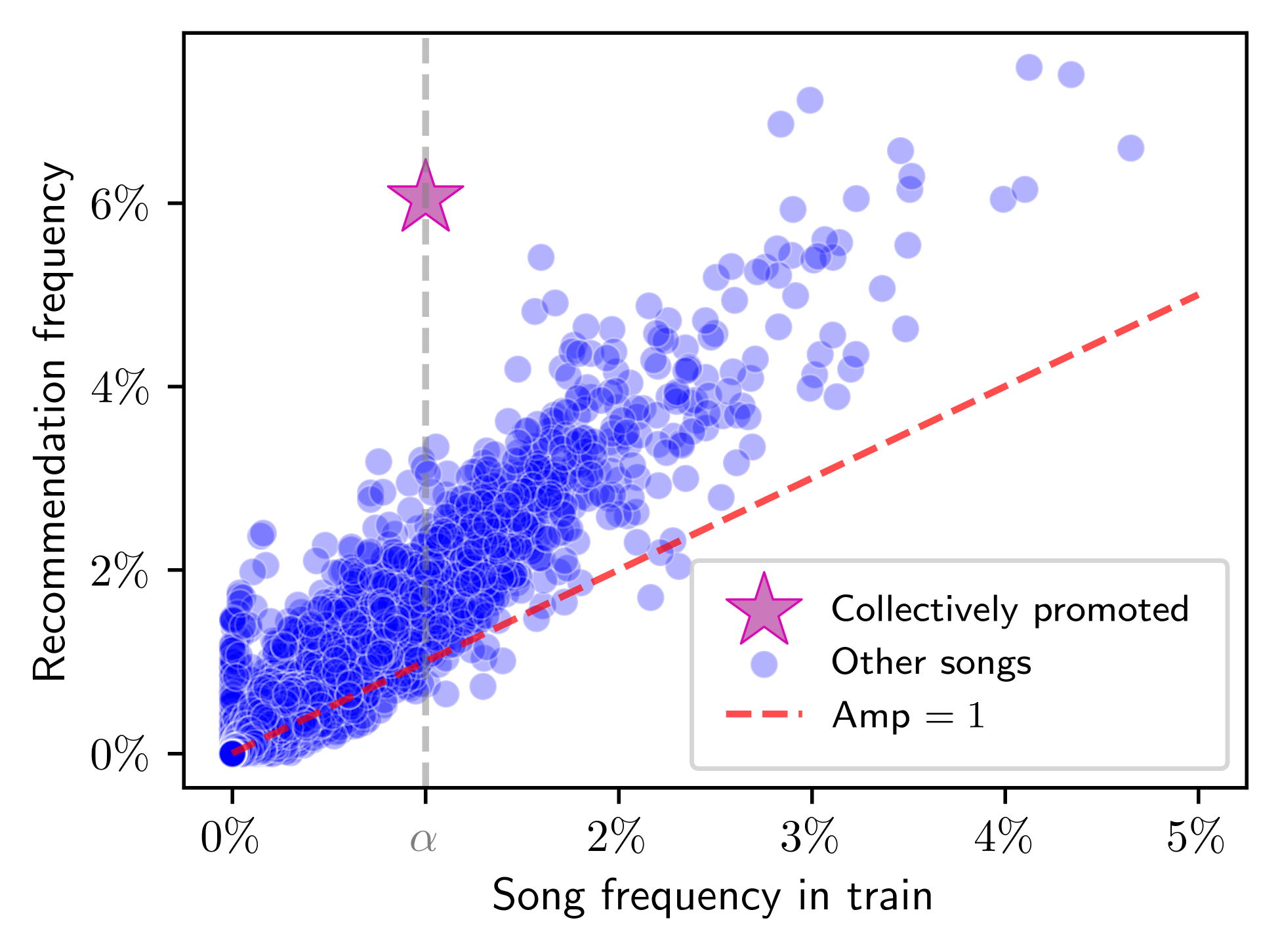}
\caption{By strategically choosing the position at which to insert the target song in a playlist, collectives can achieve a disproportionally high recommendation frequency relative to training set occurrences. The dashed red line corresponds to matching frequencies at train and test time, the blue dots correspond to naturally occurring songs.}
\label{fig:teaser}
\end{figure}

\subsection{Our work}
\label{ssec:Our_work}

We study algorithmic collective action in transformer-based recommender systems. As a case study, we consider the task of automatic playlist continuation (APC), which is at the heart of many music streaming services. APC models take a seed playlist (an ordered list of unique songs) as input and recommend songs to follow. They are trained in an unsupervised manner on the universe of playlists stored on the platform. The collective consists of platform users who can modify the subset of playlists they own. The goal of the collective is to promote a newly released target song $s^*$ by strategically placing it in their playlists.

We motivate and discuss two strategies to choose the position of $s^*$ within any given playlist. Both strategies are derived from a statistical optimality assumption on the recommender and do not require knowledge of the specifics of the model architecture or the model weights. Instead, they use that the model is trained to fit sequential patterns in existing data and build on aggregate song statistics that are feasible to gather from public information.
We empirically test our strategies using an industry-scale APC model that has been deployed to provide recommendations for millions of users on Deezer---one of the biggest streaming platforms in the world.
To train the model, we use the Spotify Million Playlist Dataset, treating each playlist as a user and randomly sampling a fraction to compose the collective.

We find that by strategically choosing the position of the target song, collectives can achieve significant over-representation at test time, see Figure~\ref{fig:teaser} for a teaser. 
We experiment with collectives composed of a random sample of users owning between $0.001\%$ and $2\%$ of the training data instances. 
Interestingly, even tiny user collectives, controlling as few as $60$ playlists, can achieve an amplification of up to $25\times$, referring to the song's recommendation frequency relative to the training frequency.
This is $40\times$ more than an average song occurring at the same frequency in the training data. In contrast, placing the song in a fixed position in every playlist is largely ineffective.

Our strategy satisfies a strict authenticity constraint on the playlists and preserves user experience at training time by design. Interestingly, we find that also at test time recommendations are largely preserved; not only on aggregate but also for members of the collective.
As a consequence, the strategies come with small externalities for users, and at the same time, they also have a relatively small effect on model performance. For large collectives controlling $>3\%$ of the playlists, the effect corresponds to every target song recommendation replacing an otherwise relevant song in less than $15\%$ of the cases, leaving other recommendations unaltered.
Lastly, we show that the newly gained recommendations are taken from artists of diverse popularity without any indication that a specific competing artist suffers disproportionally.

Taken together, our work demonstrates a first example of effective collective action in sequential recommender systems through strategic data ordering. We show how collective action goals can be achieved while largely preserving service quality and user experience. Moreover, the studied strategies are easy to implement, raising many pressing questions, challenges, and opportunities for future work.

\subsection{Related work}

The fairness of recommendation systems on online platforms remains a pressing issue for both content consumers and producers~\citep{burke2017multisided,Mehrotra2018FairMarketplace,ferraro2021_whatisfair,Ionescu2023FairnessCreators}---see~\citet{Zehlike2022FairnessinRankingII} for a detailed overview.
Beyond design choices at the platform level and organized strikes by artists, several recent works study individual strategic users attempting to influence their own recommendations~\citep{NEURIPS2018_a9a1d531,haupt2023recommending,cen2023measuring,cen2023user}. Other works consider adding antidote data to fight polarization and unfairness~\citep{Rastegarpanah2019FightingFire,Fang2022FairRoad}.

Beyond recommender systems, a related line of work focuses on optimization strategies for users interacting with machine learning systems more broadly. \citet{Vincent2021ConsciousDataContribution} call for \textit{conscious data contribution}, \citet{Vincent2019DataStrikes} discuss data strikes, and \citet{Vincent2021DataLeverage} emphasize the potential of data levers as a means to gain back power over platforms.
\citet{hardt2023algorithmic} introduce the framework of \textit{algorithmic collective action} for formally studying coordinated strategies of users against algorithmic systems. They empirically demonstrate the effectiveness of collective action in correlating a signal function with a target label to influence classification outcomes. \citet{sigg2024decline} inspect collective action at inference time in combinatorial systems. Complementing these findings, we demonstrate that collective action can be effective in generative models even without control over samples at inference time. In addition, we highlight a so far understudied dimension of algorithmic collective action by illuminating the externalities of algorithmic collective action.

At a technical level, our findings most closely relate to \textit{shilling attacks}, or more broadly, \textit{data poisoning attacks}~\cite[c.f.,][]{zhiyi23poison}.
Shilling attacks are usually realized by injecting fake user profiles and ratings in order to push the predictions of some targeted items~\citep{si2020shillingsurvey,Sundar2020UnderstandingShilling}.
Due to the fraudulent nature of these attacks, there are little design restrictions on the profiles, and they often come with considerable negative effects for the firm~\citep{Mahony2004CollaborativeRobustness,gunes2014shillingsurvey}.
Data poisoning attacks in recommender systems predominantly focus on collaborative filtering-based models, with a few exceptions;
\citet{Zhang2020PracticalDataPoisoning} propose a reinforcement learning-based framework to promote a target item, \citet{Yue2021BlackBox} provide a solution to extract a black-box model's weights through API queries to then generate fake users for promoting an item, and
\citet{Yue2022DefendingSubstitution} propose injecting fake items into seemingly real item sequences (at inference time and without retraining) with a gradient-guided algorithm, requiring full access to the model weights. Taking the perspective of collective action, we focus on easy-to-implement strategies that require minimal knowledge of the model and operate under an authenticity constraint to preserve the utility of altered playlists while seamlessly integrating into natural interaction with the platform.

Further, our work pertains to a broader scholarly literature interested in improving labor conditions for gig workers on digital platforms~\citep[e.g.,][]{jarrahi19gig, Toxtli2023}, optimizing long-term social welfare in online systems~\citep{mladenov20longterm}, and understanding dynamics in digital marketplaces~\citep{Jagadeesan_Jordan_Haghtalab_2023}. The type of strategic data modification we consider falls under the umbrella of \textit{adversarial feature feedback loops}~\citep{Pagan2023FeedbackLoops}.
Taking advantage of collective strategies to change model outcomes more broadly has been studied in tabular data~\citep{creager2023online}, computer vision~\citep{Shafahi2018PoisonFrogs}, and recently in generative AI~\citep{shan2023glaze}.

\section{Preliminaries on automatic playlist continuation}

We use automatic playlist continuation (APC) as a running example of a sequential recommendation task. APC forms the backbone of major streaming platforms, such as Spotify and Deezer.
To formally define the recommendation task, let $\cS=\{s_1, ..., s_n\}$ denote the universe of songs, where $n\geq 1$ denotes the number of unique songs. A playlist $p=[s_1,s_2,...,s_{L}]$ is composed of an ordered list of $L$ songs selected from $\cS$ without replacement. 
Given a seed playlist $p$, the firm's goal is to predict follow-up songs that the user likely listens to. We consider a top-$K$ recommender system that outputs a personalized ordered list of $K \geq 1$ songs. We write $\mathrm{Rec}_K(p)$ for the set of $K$ songs recommended for a seed playlist $p$.

\subsection{Transformer-based recommender}
\label{ssec:Transformer_based_recommender}

Over the past years, most large platforms have shifted from relying on collaborative filtering-based models for APC to building deep learning-based recommenders that account for sequential and session-based information~\citep{Hansen2020SpotifyContextualRecommendation,Tomasi2023SpotifyAPCwithRL, Moor2023ExploitingSequential}. 
A popular structure of transformer-based models is the following~\citep{Bendada2023DeezerPlaylistContinuationTransformers}: Each song $s$ is mapped to a song embedding vector $h_s=\phi(s)$, where $\phi$ denotes the embedding function. Each playlist $p$ is mapped to an embedding vector $h_p$ by aggregating the embeddings of the songs contained in the playlist as $h_{p} =g(h_{s_1},h_{s_2},...,h_{s_{L}})$, where $g$ is a sequence-aware attention function. We assume all playlists have length $L$ smaller than the attention window for the purpose of exposition. 

At inference time, for any given playlist $p$ the similarity between the seed playlist $p$ and all potential follow up songs $s\in \cS \setminus p$ is determined as
\begin{equation}
    \text{SIM}(s,p) := \langle h_s, h_p\rangle,
\end{equation}
and then, the $K$ songs with the largest similarity value are recommended in descending order of similarity. We denote the set of recommended songs as
\begin{equation}
\mathrm{Rec}_K(p)=\argmax_{S' \subseteq \cS \setminus p: |S'|=K}\; \sum_{s\in S'} \text{SIM}(s,p).
\end{equation}
The embeddings $\phi$ and the attention function $g$ are parameterized by neural networks. They are trained from existing user-generated playlists in a self-supervised manner by repeatedly splitting playlists into a seed context and a target sequence and employing a contrastive loss function for training.

\paragraph{Statistical abstraction.} In the following, we do not assume the collective has knowledge of the parameters of either $\phi$ or $g$. Instead, the design of the strategy builds on the assumption that sequential, transformer-based models are trained such that $\mathrm{SIM}(s,p)$ is large for songs $s$ that frequently follow context $p$ in the training data, and small otherwise. This approximately is robust to nuances in hyperparameter choices or architecture design and applies to any sufficiently expressive and well-trained model.

\begin{figure}[!t]
\centering
\includegraphics[width=0.7\textwidth]{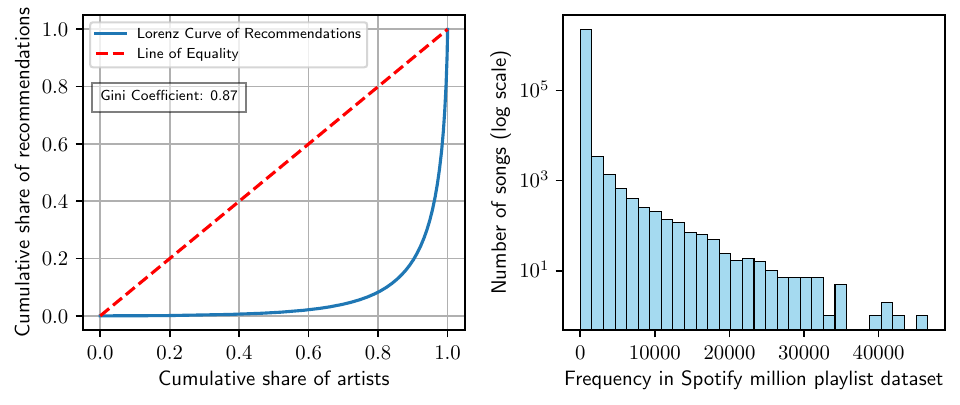}
\caption{
\textbf{Imbalance in recommendation distribution}. (left) The Lorenz curve shows that 80\% of all recommendations are concentrated among just 10\% of artists.
(right) The Spotify track frequency distribution shows the long tail of song frequencies in user-generated playlists: close to 50\% of tracks in playlists occur only once.
}
\label{fig:long_tail}
\end{figure}

\subsection{Typical imbalances in recommendations}
\label{ssec:Typical_imbalances_in_recommendations}

On today's music streaming platforms, a small number of artists receive the vast majority of recommendations, while the majority receive few or none. 
This imbalance is illustrated by the Lorenz curve in Figure~\ref{fig:long_tail}, which is based on recommendations derived from the Deezer model on the Spotify MPD dataset (see Section~\ref{sec:Experiments}).
The Gini coefficient measuring inequality corresponds to $0.87$.
Streaming and radio statistics reveal an even more severe imbalance: the top 1\% of newly released songs receive 99.99\% of radio plays and 90\% of streams go to just 1\% of artists~\citep{Blake2020data}.

Considering Figure~\ref{fig:teaser} we can also see that songs with high prevalence in the training data are recommended disproportionately often at test time compared to their training set frequency (referring to the slope of $\sim$$1.8$ of the blue point cloud). This gain in exposure through the recommender comes at the expense of many low-frequency songs that receive no recommendations at test time, further amplifying existing imbalances. Considering Spotify's substantial power to influence song consumption among platform users~\citep{Aguiar2021Platforms}, withholding initial exposure for these songs limits their potential to reach a broader audience, significantly impacting an artist's career. Thus, in this work, we focus on collective efforts to boost recommendations of a newly released song, which we add to the training data.

\section{Algorithmic collective action for promoting songs}
\label{sec:Collective_action_for_promoting_songs}
Following~\citet{hardt2023algorithmic}, we consider collectives that are composed of a fraction $\alpha\in(0,1]$ of randomly sampled users on the platform. We assume each user controls a single playlist. Let $\cP_0$ denote the distribution over playlists. Members of the collective can strategically manipulate their playlists. We use $\mu(\cdot)$ to describe the strategy of mapping an original playlist to a modified playlist.  The recommender system $\mathrm{Rec}_K$ is trained on a partially manipulated training dataset $\cD$, composed of $N$ samples from $\cP_0$, among which $\alpha N$ have been transformed under $\mu$. 

\paragraph{Success and amplification.} The goal of the collective is to increase recommendations of a target song $s^*$  for members outside the collective, i.e., for a randomly sampled playlist from $\cP_0$ at test time. We measure the success of collective action as 
\begin{equation}
    S(\alpha) := \mathrm{E}_{p\sim \cP_0}\; 1\left[s^*\in \mathrm{Rec}_K(p)\right].
\end{equation}

We are particularly interested in measuring the effectiveness of a strategy relative to the effort of the collective. Therefore, we define \textit{amplification} (Amp) as the fraction of newly gained target recommendations at test time divided by the fraction of manipulated playlists in the training set:
\begin{equation}
\mathrm{Amp}(\alpha) = \frac 1 \alpha (S(\alpha)-S(0))
\label{eq:simple_amplification}
\end{equation}
An amplification of $0$ means that the strategy is ineffective, an amplification of $1$ means that the song frequency in the training set is proportionally represented in the model's predictions, and an amplification larger than $1$ means that collective action achieves a disproportionate influence on the recommender. In the following, we choose a song $s^*$ that does not currently appear in the training data, hence $S(0)=0$.

\subsection{Authenticity constraint}
\label{sec:Coordinateditempromotion}

Participants of the collective are users of the platform. We design collective action strategies under the following authenticity constraint, to ensure minimal impact on user experience:
\begin{definition}[Authenticity constraint] We say a strategy $\mu:p\rightarrow p'$  is authentic iff the Levenshtein distance between $p$ and $p'=\mu(p)$ satisfies $\mathrm{Lev}(p,p')\leq 1$ for any $p$.
\end{definition}

The Levhenstein distance~\citep{levenshtein1966binary}, also known as edit distance in information theory, counts the number of operations needed to transform one sequence into another. The song insertion strategy we propose in this work is one concrete instantiation of $\mu$ that satisfies this constraint. More specifically, our strategy consists of inserting an agreed-upon target song $s^*$ at a specific position in every playlist $p$, corresponding to edit distance $1$. Existing adversarial strategies typically perform larger modifications to playlists and would not satisfy this constraint~\citep{Zhang2020PracticalDataPoisoning,Yue2021BlackBox}.

\subsection{Algorithmic lever}
\label{sec:Algorithmic_lever}
For our song insertion strategy the collective strategically chooses, for each playlist $p$, the position $i^*$ at which to insert the target song $s^*$. Let $p_{i^*}^-$ denote the sequence of songs $[s_1,...s_j]$ in $p$ up to index $j=i^*$. 
Given the sequential nature of the recommender, strategically placing $s^*$ at position $i^*$ implies that  the similarity between seed context $p_{i^*}^-$ and $s^*$ is increased. Thus, by choosing a specific index $i^*$, the collective can target very specific contexts.

\begin{figure}
\floatname{algorithm}{}
\begin{algorithm}[H]
\caption*{Strategies: {\textcolor{DarkGreen}{(a) \texttt{InClust}}} and {\textcolor{DarkRed}{(b) \texttt{DirLoF}}}}
\label{pseudocode:DirLoF}
\begin{algorithmic}[1]
\Require $s^*$, collectively owned playlists $D^*=\{p_1, p_2, ..., p_{n}\}\subseteq D $
\State \textbf{Coordination step:}
\State \textcolor{DarkGreen}{(a) $r^s\leftarrow$ for every $s$ in $D^*$ pool information to count song frequencies in $D^*$.}
\State \textcolor{DarkRed}{(b) $q^s\leftarrow$ for every $s$ in $D^*$ estimate training set song frequency by gathering side information.}
\For{all playlists $p \in D^*$}
    \State \textbf{Define anchor $s_0$:} find song $s_0 \in p$ such that \textcolor{DarkGreen}{(a) $r^{s_0}\geq r^s\; \forall s\in p$ } or \textcolor{DarkRed}{(b) $q^{s_0}\leq q^s\; \forall s\in p$}
    \State \textbf{Insert target song:} insert $s^*$ {\textcolor{DarkGreen}{(a) before $s_0$}} or {\textcolor{DarkRed}{(b) after $s_0$}}
    \State Store modified playlist
\EndFor
\end{algorithmic}
\end{algorithm}
\centering
\begin{subfigure}[!htbp]{0.45\textwidth}
\centering
\includegraphics[width=0.7\textwidth]{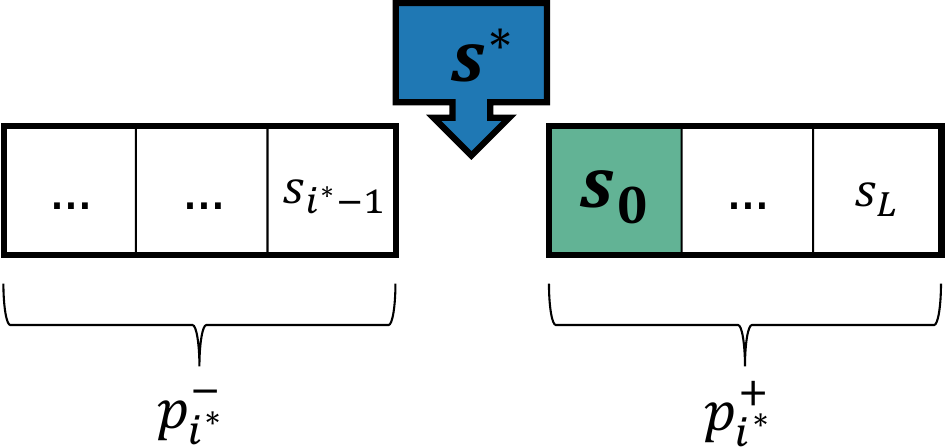}
\captionsetup{
    labelfont={color=DarkGreen},
    textfont={color=DarkGreen}
}
\caption{\texttt{InClust}}
\label{fig:anchors_indirect}
\end{subfigure}
\hspace{5mm}
\begin{subfigure}[!h]{0.45\textwidth}
\centering
\includegraphics[width=0.7\textwidth]{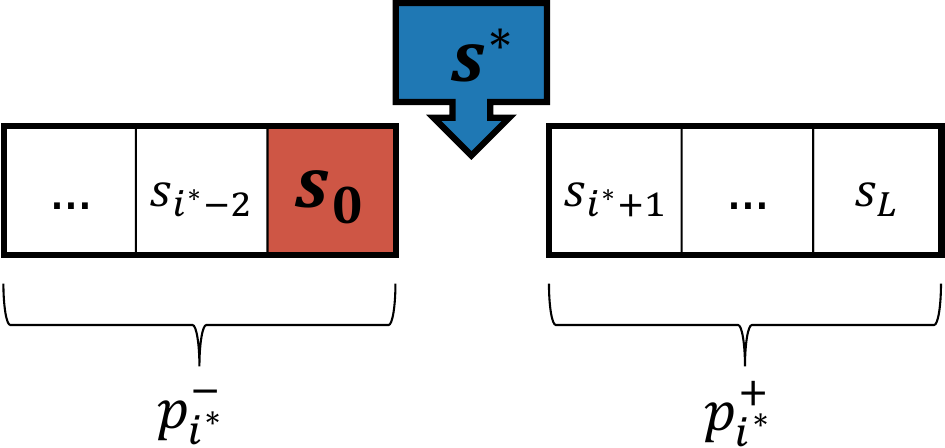}
\captionsetup{
    labelfont={color=DarkRed},
    textfont={color=DarkRed}
}
\caption{\texttt{DirLoF}}
\label{fig:anchors_direct}
\end{subfigure}
\caption{Song insertion strategies, pseudocode and illustration.}
\label{fig:pseudocode}
\end{figure}

Inclusion in the set $\mathrm{Rec}_K(p)$ leads to a song's recommendation for context $p$ at test time. In turn, being ranked in position $K+1$ does not yield any recommendations. Instead, the probability mass in the tails is reallocated to the top $K$ songs. 
To effectively boost recommendations of $s^*$ at test time, the song $s^*$ needs to be among the top $K$ songs with high frequency over a randomly sampled context $p$. 
Our strategies exploit two different algorithmic levers towards this goal:

\paragraph{Concentrating effort.} 
Members of the collective can target specific high-leverage contexts in a coordinated fashion to increase the likelihood of inclusion. Compared to random song placement, the collective can increase the mass on a particular context by a factor of $L$. Thus, coordination can pay off disproportionally if additional probability mass is sufficient to reach the top $K$ threshold. The {\textcolor{DarkGreen}{\texttt{InClust}}} strategy projects this intuition from the non-parametric setting to the embedding space of the recommender. It implements a way for selecting similar contexts in embedding space. Namely, it systematically places $s^*$ directly \emph{before} each occurrence of a popular song $s_0$. Thereby it \underline{in}directly targets \underline{Clust}ers of similar contexts around $h_{s_0}$ in embedding space. To implement this strategy, the collective repeatedly determines the most frequent song in their playlists, places $s^*$ before every occurrence of this song, and then repeats this with the remaining playlists until all of them are used. Pseudocode can be found in Figure~\ref{fig:pseudocode}.

\paragraph{Strategically exploiting overrepresentation.}
An alternative lever the collective has available is to strategically target contexts that are overrepresented among the playlists the collective controls. Meaning that the frequency of the context among the playlists owned by the collective is larger than the overall frequency in the training data due to finite sample artifacts. The {\textcolor{DarkRed}{\texttt{DirLoF}}} strategy aims to identify such contexts by \underline{Dir}ectly targeting \underline{Lo}w-\underline{F}requency contexts. Thereby it exploits the long-tail nature of the overall song frequency distribution (see Figure~\ref{fig:long_tail}). The core intuition is that if they manage to target low-frequency contexts, a single song placement might be sufficient to overpower existing signals. To identify low-frequency contexts, the collective uses the frequency of the last song as a proxy. For each playlist, it selects the anchor songs $s_0$ with the smallest overall song frequency and places $s^*$ right after $s_0$.

\subsection{Obtaining song statistics}
Our strategies rely on statistics about the occurrence of individual songs in the training data. 
The \texttt{InClust} strategy targets contexts preceding a high-frequency song, whereas the \texttt{DirLoF} strategy targets contexts ending on a low-frequency song. In the implementation of the two strategies there is an important difference. \texttt{InClust} can be implemented from only statistics obtained from the playlists the collective owns; all it requires is participants to set up infrastructure for pooling and counting songs in their playlists, e.g., through an app or an online service. In contrast, to effectively implement the \texttt{DirLoF} strategy, the collective needs statistics about the full training data to identify the songs that are least popular overall.
However, they typically do not have direct access to this information and need to estimate proxies using other data sources instead. For example, they can use publicly available user-generated playlists, which are often accessible through official APIs. Additionally, scraping external data sources can provide supplementary information. We implement and evaluate the use of scraped song streams for \texttt{DirLoF} in Section~\ref{ssec:DirLoF_with_proxy}.~\looseness=-1

\section{Empirical evaluation}
\label{sec:Experiments}

We evaluate\footnote{
The code is available at \codeurl.} our collective action strategies against a public version of Deezer's transformers-based APC solution that ``has been deployed to all users''~\citep[p.~472]{Bendada2023DeezerPlaylistContinuationTransformers}.
To train the model, we use the Spotify Million Playlist Dataset (MPD), which is currently the largest public dataset for APC~\citep{SpotifyRecSysChallenge2018}.
It contains one million playlists generated by US Spotify users between 2010 and 2017, with an average length of 66.35 tracks from nearly 300,000 unique artists.

\paragraph{Model training and evaluation.}
\label{ssec:Methodology}

We use the standard methodologies used in APC for model training and testing.
We randomly select 20,000 playlists to build a test and validation set of equal sizes. The remaining 980,000 playlists are used for training the model. The collective intervenes by strategically modifying an $\alpha$ fraction of the playlists composing the training and validation set.
We consider collectives of size $\alpha\in[0.00001, 0.02]$ which corresponds to $10$ to $20000$ playlists.
For evaluation on the test set, every playlist $p$ is split into a seed context and a masked end. The length of the seed context is chosen randomly in $[1,10]$ for each playlist and models are evaluated by comparing the model's recommendations based on the seed playlist to the masked ground truth.
We employ five-fold cross-validation, using different random seeds for sampling the playlists designated for training, validation, and testing, as well as for selecting the subset controlled by the collective. We use bootstrapped 95\% confidence intervals (CI) over folds when reporting results.

\paragraph{Baselines.}
We consider various baseline strategies to compare with our collective strategies, each performing the same number of target song insertions. The \texttt{Random} strategy inserts $s^*$ at a \texttt{Random} position in the playlist and the \texttt{AtTheEnd} strategy places $s^*$ as the last song of the playlist. In addition, \texttt{Insert@$i$} inserts the song always at position $i$ in every playlist, and \texttt{Random@$i$-$j$} inserts $s^*$ at a random position between indices $i$ and $j$. Unlike our collective strategies, these baselines do not require coordination among participants beyond the shared goal of promoting $s^*$.

\begin{figure*}[t!]
\centering
\includegraphics[width=0.65\textwidth]{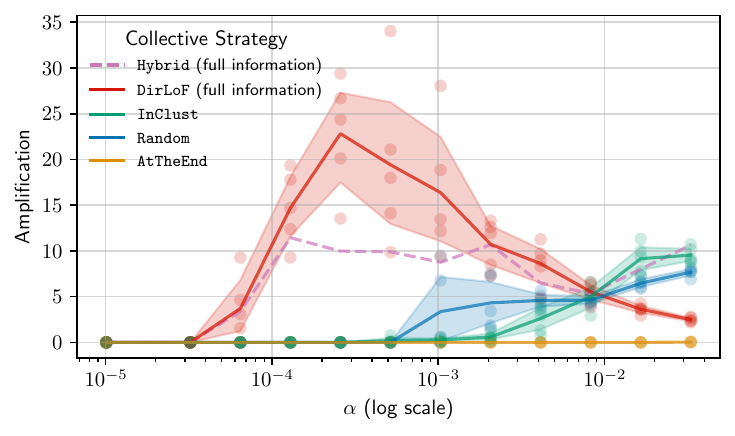}
\caption{\textbf{Success of our collective action strategies.} For tiny collectives \texttt{DirLoF} achieves an amplification of up to $25\times$ while uncoordinated strategies (\texttt{Random}, \texttt{AtTheEnd}) are mostly ineffective. For larger collectives, \texttt{InClust} outperforms \texttt{DirLoF}. Amplification significantly exceeds $1$, implying a disproportional test-time effect due to targeted song placement.
}
\label{fig:empirical_success_and_amplifications}
\end{figure*}

\subsection{Success of collective action}
\label{ssec:Empirical_success}

We start by evaluating the success of the proposed strategies, assuming full information about song frequencies in the training data. 
In Figure~\ref{fig:empirical_success_and_amplifications}, we plot the amplification for different $\alpha$. Recall that amplification measures how much more frequently the target song is recommended at test time compared to its training set frequency. Since all strategies insert the same number of target songs into the training data, amplification directly translates into relative recommendation frequency.

First, observe that strategic song placement allows very small collectives ($\alpha\leq 0.1\%$) to be successful, whereas \texttt{Random} or fixed placement of $s^*$ is largely ineffective.
For $\alpha=0.025\%$, the \texttt{DirLoF} strategy achieves amplification of up to $25$. Zooming into Figure~\ref{fig:teaser}, we can see that in contrast to an average song that naturally occurs with the same frequency in the training data, the number of recommendations is $40\times$ larger. This suggests that collective action could make a tremendous difference for these artists: suppose an artist's song is streamed 10,000 times, yielding a revenue of $\$40$ at a royalty rate of $\$0.004$ per stream~\citep{Marshall2025Let}; an amplification of $25$ would hypothetically increase this revenue to $\$1,000$. While this example is purely illustrative (as actual royalties depend on the platform and payment model used), it emphasizes the link between recommendations and potential revenue.

For collective sizes of $\alpha\geq 0.1\%$ the \texttt{InClust} starts being effective, as it has enough mass to  compete with existing signals associated with a cluster of similar context embeddings. As the strategy can target several such clusters at the same time, amplification increases with $\alpha$ though with diminishing returns, achieving $\text{Amp}=10$ for $\alpha \approx 2\%$. From Figure~\ref{fig:teaser}, we can see that in the regime of $2\%$ training data frequency, a typical song enjoys an amplification of $1.8$, which is more than a factor of $5$ smaller. Additional ablations to study the inner workings of the \texttt{InClust} strategy are presented in Appendix~\ref{ssec:AblationStudyIndirectContextEmbeddingAttacks}.

We also observe that the success of the random strategy increases with the collective size. This implies that even minimal coordination, in which members agree to all insert the same song $s^*$, \emph{independent} of the playlist they own, can already lead to significant amplification.
Amplification values for the other baselines inserting $s^*$ at a fixed position are all close to $0$, see Table~\ref{rebuttal:new_baselines}.

\begin{table}[t!]
    \centering
    \caption{\textbf{Placing song at fixed position is largely ineffective.}
    Mean Amplification (Std Dev) for insertion of $s^*$ at a fixed index $i$.
    The best competing strategy is highlighted in bold.
    }
    \label{rebuttal:new_baselines}
    \small{
    \begin{tabular}{llll}
        \toprule
        Strategy & $\alpha=0.0002$ & $\alpha=0.001$ & $\alpha=0.002$ \\
        \midrule
        \texttt{Insert@$0$} & 0.00 (0.00) & 0.00 (0.00) & 0.00 (0.00) \\
        \texttt{Insert@$1$} & 0.00 (0.00) & 0.00 (0.00) & 0.00 (0.00) \\
        \texttt{Insert@$3$} & 0.00 (0.00) & 0.00 (0.00) & 0.00 (0.00) \\
        \texttt{Insert@$5$} & 0.00 (0.00) & 0.00 (0.00) & 0.00 (0.00) \\
        \texttt{Insert@$7$} & 0.00 (0.00) & 0.02 (0.04) & 0.07 (0.08) \\
        \texttt{Random} & 0.00 (0.00) & \textbf{3.36 (4.45)} & \textbf{4.32 (2.90)} \\
        {\textcolor{DarkRed}{\texttt{DirLoF}}} & {\textcolor{DarkRed}{22.82 (6.21)}} & {\textcolor{DarkRed}{16.38 (7.37)}} & {\textcolor{DarkRed}{10.73 (2.66)}} \\
        \bottomrule
    \end{tabular}}
\end{table}

\paragraph{\texttt{Hybrid} strategy.} Building on the empirical insights from this section we construct a hybrid strategy that interleaves the two approaches by first using \texttt{InClust} to target indirect anchors that appear at least $\lambda$ times in the collective and then switches to \texttt{DirLoF} for playlists where no such anchor is present (we use $\lambda=10$). This corresponds to the dashed line in Figure~\ref{fig:empirical_success_and_amplifications}. We will come back to this strategy in Section~\ref{ssec:Internalities_and_externalities}.

\paragraph{Robustness to hyperparameters.}
Recall that our strategies are designed based on a statistical intuition of sequential generation and do not require any knowledge of the model parameters. Thus their performance is insensitive to details of the model architecture. We demonstrate the robustness with additional experiments where we vary the hyperparameters of the model, see Table~\ref{rebuttal:hp_robustness} in Appendix~\ref{app:robustness}. However, at the same time, the design of our strategies relies on the assumption that the model approximates the conditional probabilities in the training data sufficiently well. Accordingly, the effectiveness of the strategy decreases if model training is stopped early, as demonstrated in Table~\ref{rebuttal:nr_of_epochs} in Appendix~\ref{app:robustness}.

\begin{figure}
    \centering
\includegraphics[width=0.75\textwidth]{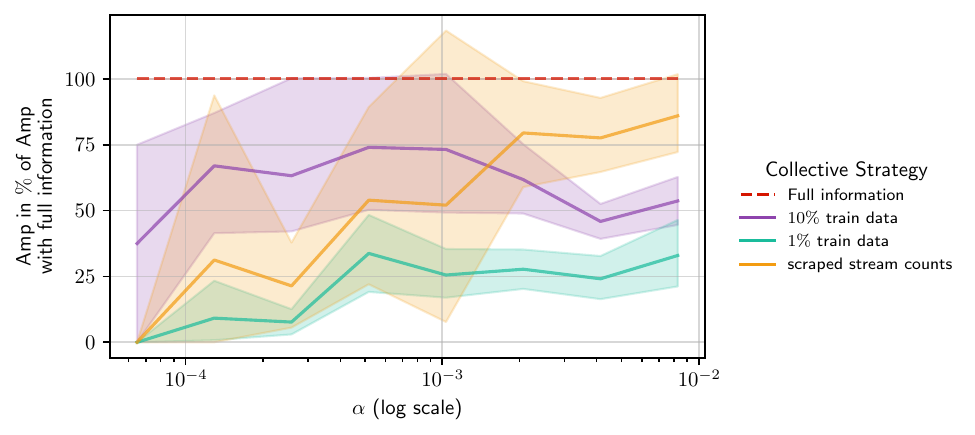}
    \caption{
    \textbf{Information bottleneck.} The empirical amplification of the \texttt{DirLoF} strategy decreases with worse song statistics but scraped song streaming counts can serve as a practical solution.}
    \label{fig:empirical_amplification_partial_information_frac}
\end{figure}

\subsection{\texttt{DirLoF} strategy with approximate song statistics}
\label{ssec:DirLoF_with_proxy}

The \texttt{DirLoF} strategy critically relies on training data song frequency estimates to determine the low-frequency anchor songs. We investigate the strategy's success with partially available song information in Figure~\ref{fig:empirical_amplification_partial_information_frac}. We find that if a collective of size $\alpha = 1\%$ has access to an additional $1\%$ of the remaining training data, they can already achieve $\approx 30\%$ of the amplification in the full information setting, with $10\%$ of the data, it is $>50\%$ of the achievable amplification.

By default, user-generated playlists on streaming platforms are often publicly accessible, enabling researchers to gather song frequency data through API calls.
However, the amount of training data that can be aggregated is limited by the platform's API rate limits.
Alternatively, proxy statistics can be used to increase the fraction of songs for which estimates are available.
To illustrate the feasibility of this approach, we implemented a scraper to obtain current stream counts from Spotify.
Although these counts are visible in the Spotify browser version, they are not accessible through the Spotify API.
The scraped data reflects song popularity as of 2024, which is not ideal given training is performed on the much older Spotify MPD dataset (collected between 2010 and 2017).
Nonetheless, these counts serve as effective proxies, as Figure~\ref{fig:empirical_amplification_partial_information_frac} impressively shows.~\looseness=-1

Despite the temporal gap, a collective of size $\alpha=1\%$
can achieve over $85\%$ of the amplification achievable in a full-information setting simply by using scraped 2024 stream data to approximate past popularity levels.
Even a smaller collective of $\alpha=0.1\%$ can reach about $50\%$ of the amplification seen in the full-information scenario.
In practice, scraped stream counts are likely to be more accurate proxies, as models are typically trained on more recent data. However, within the scope of our study, it remains impossible to access historical stream counts that would reflect popularity as of the time the playlists were originally generated. Thus our proof of concept should be seen as a lower bound on the quality of the statistics that can be extracted from external information.

\begin{figure}[t!]
\centering
\includegraphics[width=0.98\textwidth]{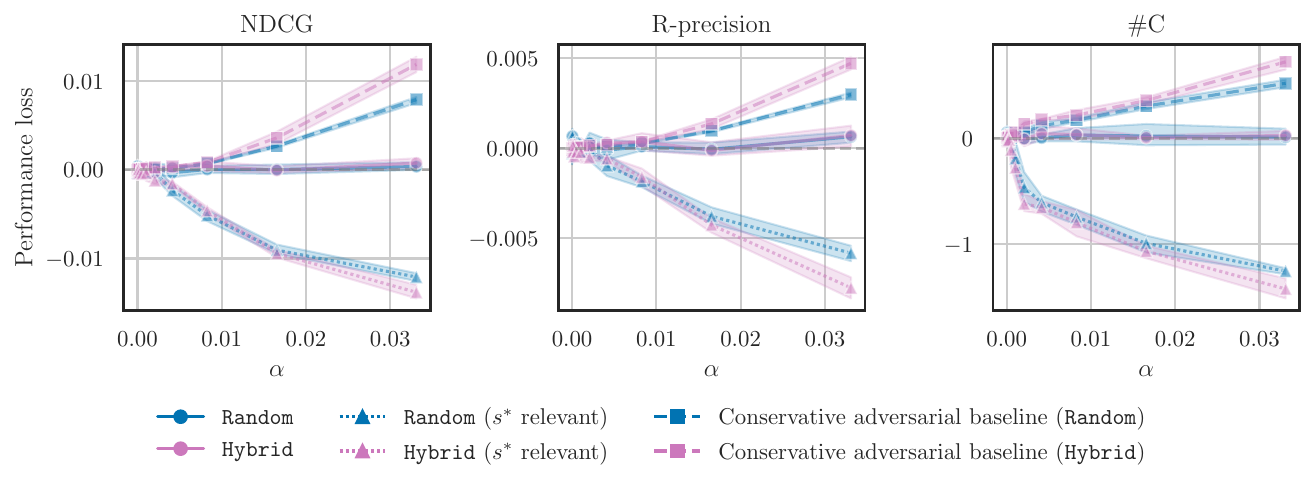}
\caption{Effect of algorithmic collective action on recommendation performance. Performance loss relative to training on clean data for the \texttt{hybrid} / \texttt{random} strategies (solid lines), a conservative adversarial baseline (dashed lines), and an optimistic scenario where $s^*$ is treated as relevant (dotted lines).
}
\label{fig:effect_on_performance_NDCG_only}
\end{figure}

\subsection{Internalities and externalities of algorithmic collective action}
\label{ssec:Internalities_and_externalities}

We now inspect the effect of our strategies on different stakeholders in the system, including the firm, other artists, and the members of the collective themselves. For this investigation, we focus on the hybrid strategy and contrast it with a conservative adversarial baseline.

First, we gauge the impact of collective action on the firm. This helps us understand the overall quality degradation of the service and the incentives of the firm to protect against collective action. We compare the performance under a recommender trained on the clean data and a recommender trained on the manipulated data. Figure~\ref{fig:effect_on_performance_NDCG_only} shows the corresponding loss in performance due to collective action for three different evaluation metrics. We find that our strategy (solid lines) has only a marginal effect on the recommender's performance. For comparison, we also present a conservative adversarial baseline (dashed lines). This baseline achieves the same number of target song recommendations, but simulates a scenario where the target song recommendation results in the first relevant item in playlist recommendations being replaced. The considerably larger performance loss of this baseline indicates that our strategy only rarely affects relevant songs. 
Finally, as a thought experiment, consider $s^*$ as a relevant recommendation (dotted lines). The fact that collective action enhances the system's performance, means that the overall fraction of compromised recommendations due to the intervention is much smaller than $\alpha$ for any level of $\alpha$.

\begin{figure}[t]
    \centering
    \includegraphics[width=0.75\textwidth]{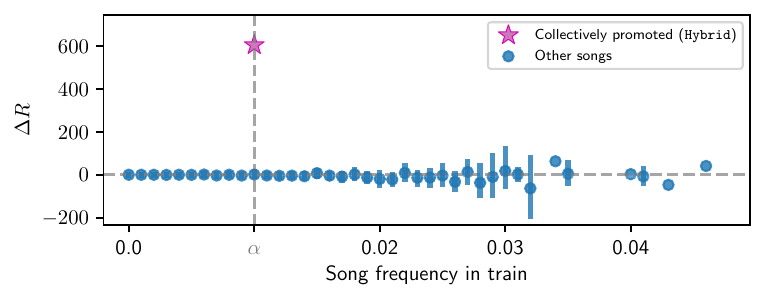}
    \caption{Impact of collective action on other songs. We use $\alpha=1\%$ to obtain an upper bound on the effect. $\Delta R$ denotes the change in the number of recommendations for a song due to collective action. Songs are sorted by their training set frequency and aggregated into 50 evenly spaced bins, whose means are represented by the blue dots with 95\% CI.}
    \label{fig:impact_on_other_artists}
\end{figure}

Second, we inspect the effect of collective action on other artists. To this end, Figure~\ref{fig:impact_on_other_artists} depicts the change in recommendations for individual songs of different popularity levels. Songs are binned by frequency and the bars indicate variation across songs. The star shows the target song $s^*$, which, as intended, is significantly more affected by collective action than any other song. We also see that recommendations replaced by the target song seem to span songs of all popularity levels. There is no indication that our strategy would harm specific songs or artists disproportionally. Also no other song is impacted nearly as much as the target song.~\looseness=-1

Finally, we focus on the experience for participants who listen to the playlists. At training time our strategies are designed to only ask for minimal modifications with the goal to preserve user experience for members of the collective. We envision this to be an important factor for incentivizing participation in practice.
Non-participating individuals are not affected at this stage. At test time, we have seen in Figure~\ref{fig:effect_on_performance_NDCG_only} that recommendation performance is largely preserved. More precisely, participating in collective action does not deteriorate the fraction of relevant songs participants get recommended, i.e., performance remains equivalent across all three recommendation quality metrics (see Figure~\ref{fig:ablation_prediction_stability} in Appendix~\ref{app:EffectonArtists}). 

Taken together the results show that, compared to an adversarial approach that achieves the same payoff for the collective, our strategy is more targeted and comes with smaller externalities.

\section{Conclusion}
\label{sec:discussion}

This work studies how collective action strategies can empower participants to exert a targeted influence on platform-deployed algorithms. By experimenting with an industry-scale transformer-based APC model, we demonstrate how strategically inserting a single song within randomly sampled playlists in the training data, can much more effectively increase recommendations of that song compared to random or fixed placement.
Intriguingly, the strategies only rely on a statistical optimality assumption of the model rather than specifics of its instantiation. 

The concept of participating in collective action to steer recommendations is grounded in the idea that users on online platforms should leave their digital traces more consciously. Thereby, their consumption behavior functions as a lever to reclaim some control over the data that platforms use to predict and recommend future content.
Our emphasis on authenticity is important for preserving user experience while interacting with the platform. Similarly, the fact that the strategy only minimally interferes with service quality ensures that recommendations for other users on the platform are largely preserved.

The true power of algorithmic collective action lies in mobilizing a sufficiently large number of participants around a shared objective. This allows underrepresented artists to gain visibility through coordination. In our case, coordination corresponds to agreeing on a target song and an insertion procedure. The actual implementation of the strategy is possible with very limited technical skills and knowledge of the algorithm. We demonstrate how information for setting the parameters of the strategy can effectively be gathered using web scraping techniques. What we leave for future work is the actual implementation of an app to orchestrate collective action and share all the relevant information with the participants.

Our work suggests a widely unexplored design space for effective collective action strategies that differ from typical adversarial data poisoning attacks~\cite[c.f.][]{zhiyi23poison,Zhang2020PracticalDataPoisoning,Yue2021BlackBox,Yue2022DefendingSubstitution}. They can offer a powerful data lever to counter existing power imbalances~\citep{Vincent2019DataStrikes, Vincent2021DataLeverage}, and a community-centric approach to participatory AI~\citep{abeba22participatoryAI}. At the same, time Figure~\ref{fig:teaser} suggests that similar strategies are not yet being used, at least not during time of data collection. Understanding the role of economic power~\cite{hardt2022performative, hardt2023performativePastandFuture}, formalizing incentives~\cite{olson1965logic}, investigating protection strategies for the firm~\cite{steinhardt17protect} as well as quantifying long-term payoffs, dynamics, and equilibria, under collective action promises to be a fruitful direction for future work.

\section{Limitations and potential for misuse}
\label{app:Limitations}
Grounding algorithmic collective action means identifying both its opportunities and challenges.
The power that arises from gaining control over the learning algorithm through collective action can also be abused by individuals controlling a substantial number of playlists. Instead of collective goals, these individuals could leverage similar methods to pursue individualistic goals, creating a different incentive structure and potentially posing a risk to the system. Similarly, popular artists could use our strategy to gain additional exposure and reinforce inequalities among artists. Thus, incentive structures will crucially determine the desirability of the resulting market outcome. Designing larger-scale collective action strategies that promote fairness and equity on online platforms as well as mechanisms that disincentivize malicious use remains a crucial open question.

\section*{Acknowledgements}
We would like to thank Moritz Hardt for many formative discussions throughout the course of this project.
We would also like to thank 
Mila Gorecki, Ricardo Dominguez-Olmedo, Ana-Andreea Stoica and André Cruz for invaluable feedback on the manuscript,
and Olawale Salaudeen, Florian Dorner, Stefania Ionescu and Tijana Zrnic
for helpful feedback on earlier versions of this work. Celestine Mendler-Dünner acknowledges financial support from the Hector foundation.

\bibliographystyle{unsrtnat}
\bibliography{references}

\appendix
\newpage

\section{Song recommendation inequality}

\begin{figure}[h]
\centering
\includegraphics[width=0.43\textwidth]{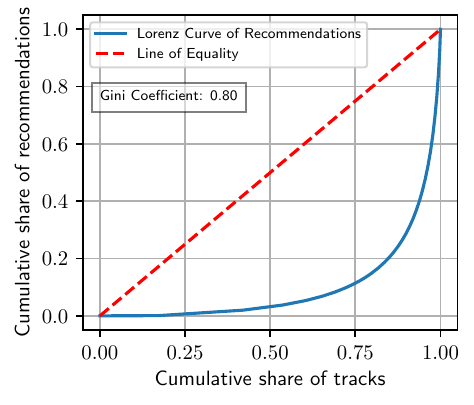}
\caption{
The Lorenz curve shows the unequal distribution of recommendations across tracks: 80\% of all recommendations are concentrated among just 15\% of tracks.
}
\label{fig:lorentz_curve_track}
\end{figure}

Figure~\ref{fig:lorentz_curve_track} visualizes the track-level distribution of algorithmic exposure with the cumulative share of recommendations (y-axis) plotted against the percentiles of tracks (x-axis).
The recommendations are derived from the Deezer model~\citep{Bendada2023DeezerPlaylistContinuationTransformers} on the Spotify MPD dataset~\citep{SpotifyRecSysChallenge2018}.
More precisely, they are based on the outputs generated for a random selection of 10,000 seed playlists for testing, produced by a model that has been trained on the remainder of the dataset, without any collective action---see Section~\ref{sec:Experiments} for more details.
Similar to the artist-based Lorenz curve in Figure~\ref{fig:long_tail}, we observe a very high level of inequality with a Gini coefficient of $0.8$ (measuring the gap between the line of equality and the Lorenz curve).

\section{Experimental details}
\label{app:ExperimentRunDetails}

All experiments were run as jobs submitted to a centralized cluster, using the open-source \texttt{HTCondor} job scheduler~\citep{HTCondor}.
For all jobs for data preprocessing and performing the data modifications as per a strategic collective action, 1 CPU was used with an allocated 100GB of RAM.
In a subsequent step, transformer models were trained using a single NVIDIA A100-SXM4-80GB GPU.
For each job, data preprocessing takes roughly 1-2 hours to complete (with coordinated strategies taking longer than uncoordinated ones).
Models are trained for 18 epochs, using the optimal hyperparameters provided by~\citet{Bendada2023DeezerPlaylistContinuationTransformers}, which takes roughly 6 hours.

A total of 1195 experiments were run:\footnote{The complete code is available at \codeurl.} We investigated 10 strategies (including 6 \texttt{DirLoF} strategies with varying levels of song statistics knowledge and 8 simple baselines, as well as 6 different hyperparameter configurations), across 12 collective sizes ($\alpha$), and an additional baseline without collective action ($\alpha=0$) as a reference for the main experiments. For the ablation study, 6 strategies were tested over 13 $\alpha$ values. Each experiment was conducted with five folds using different random seeds. This resulted in approximately 2390 CPU hours and 7170 GPU hours of total compute usage. ~\looseness=-1

\section{Additional experiments}

\subsection{Ablation study for \texttt{InClust} strategy}
\label{ssec:AblationStudyIndirectContextEmbeddingAttacks}

\begin{figure}[t!]
\centering
\includegraphics[width=0.9\textwidth]{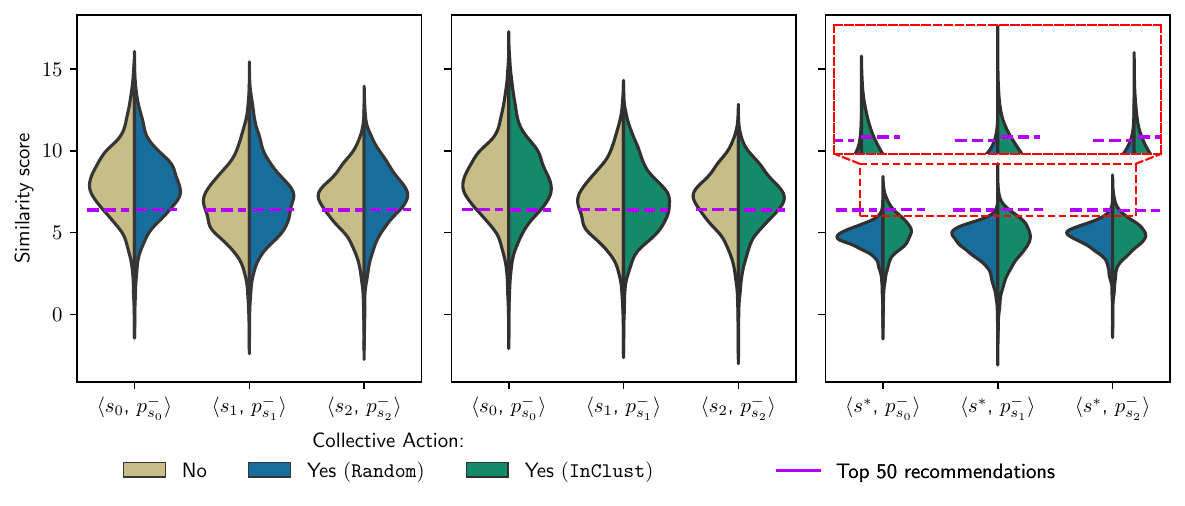}
\caption{Similarities of context embeddings for indirect anchor songs ($s_0$, $s_1$, and $s_2$) and the target song ($s^*$) for $\alpha = 1.1\%$. Dashed purple lines represent average thresholds to be among the top 50 most similar songs.
}
\label{fig:ablation_context_similarities}
\end{figure}

We perform an additional empirical investigation to provide insights into the inner workings of the strategy. In particular, the effect of strategic positioning using indirect anchors on the attention function. Recall that the recommender is trained such that $\mathrm{SIM}(s,p)= \left\langle h_{s_0}, g\left(p\right)\right\rangle$ is large for songs $s$ that frequently follow context $p$ in the training data, and small otherwise. The embedding function $\phi$ (described in Section~\ref{ssec:Transformer_based_recommender}) is insensitive to the song ordering within playlists, and strategic positioning will only surface on the attention function $g$.

For illustration, we use three indirect anchors corresponding to the three songs that occur most frequently in the playlists of the collective, denoted $s_0$, $s_1$, and $s_2$. We compare \texttt{InClust}, where we insert the target song $s^*$ before these songs, with \texttt{Random}, where we add $s^*$ in the same playlists but at random positions. This ensures we have the same pretrained song embeddings across the two strategies. Playlists that do not contain any of those three songs are not manipulated. We use $p_{s_i^-}$ to denote the context targeted by using $s_i$ as an indirect anchor.

In Figure~\ref{fig:ablation_context_similarities}, we visualize the similarity scores of the songs ($s^*$, $s_0$, $s_1$, and $s_2$) with the three context clusters that have been targeted. More specifically, for every anchor song $s_i$, we use all seed contexts $p_{s_i^-}$ that have been targeted in the training data and compare the similarity score of these contexts with (yes) and without (no) collective action. This results in a distribution over scores, as visualized by the violin plot.
The left and the middle panel show that the similarity scores of the anchor songs and their associated contexts are not altered for either of the two strategies (\texttt{Random} and \texttt{InClust}).
Furthermore, as can be seen in the right panel, the \texttt{InCLust} strategy is very effective in getting the target song to be among the top 50 most similar tracks for the targeted contexts (corresponding to the probability mass above the purple threshold). 
In contrast, for the random placement that does not specifically target these contexts, $s^*$ generally fails to achieve a ranking in the top 50.

\begin{figure*}[t]
\centering
\includegraphics[width=0.5\textwidth]{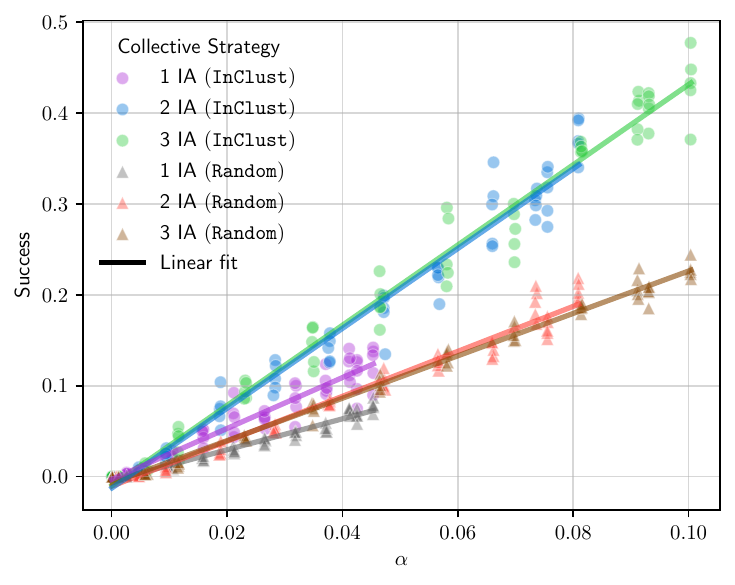}
\caption{Success with respect to the number of used indirect anchors (IA), in random or coordinated fashion.
Each dot or triangle corresponds to a separate training run.
}
\label{fig:ablation_success}
\end{figure*}

\paragraph{Targetting multiple context clusters.}
The \texttt{InCLust} strategy is effective on all three distinct context clusters that are targeted simultaneously with a single target song $s^*$.
We confirm the effectiveness of this strategy by assessing its success with respect to a test set, which contains a randomly drawn set of playlists (each split into a seed context and a ground truth) that have not been seen during training.
Figure~\ref{fig:ablation_success} shows that for any number of indirect anchors, the \texttt{InClust} strategy significantly outperforms the \texttt{Random} placement strategy.
Furthermore, it also clearly shows that it is possible to effectively target multiple context clusters using the same target song.
In conclusion, it is possible to compete with several context clusters simultaneously.
\\\\
Overall, this ablation indicates that inserting a single song in a subset of playlists can be effective in associating $s^*$ with specific contexts while preserving recommendations for songs $\cS\backslash s^*$.

\subsection{Information bottleneck}

\begin{figure*}[!htb]
\centering
\includegraphics[width=1\textwidth]{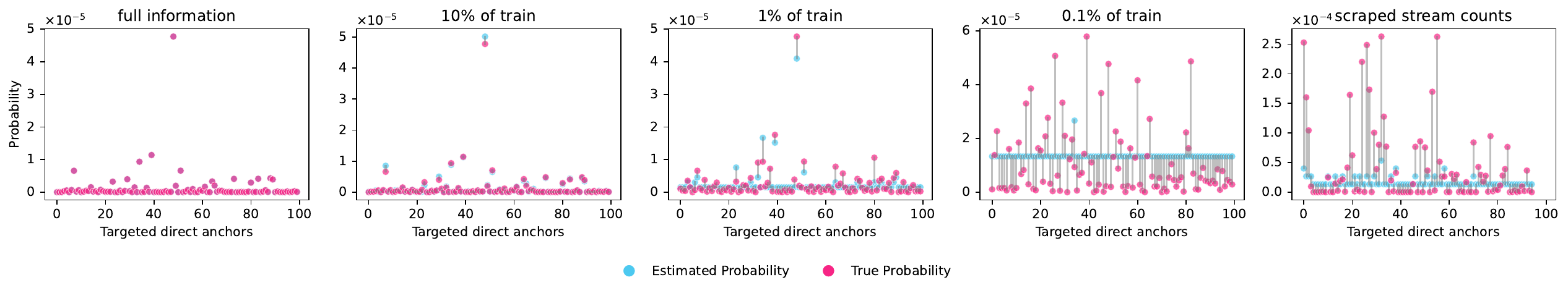}
\caption{
Estimated and true probabilities of targeted direct anchors with limited information about training set frequencies ($\alpha = 0.01\%$).
}
\label{fig:estimates_vs_true_probability}
\end{figure*}

Figure~\ref{fig:estimates_vs_true_probability} displays the 100 targeted direct anchors for $\alpha=0.01\%$ under different levels of knowledge about the song frequencies in the training set.
Less information results in the selection of more frequent songs, as the gap between the estimated probability (relative occurrences of songs in the known fraction of the dataset or estimated using external information) and the true probability (relative occurrences of songs in the entire training data) widens.
This difference provides insight into the reduced amplification observed in Figure~\ref{fig:empirical_amplification_partial_information_frac} for a specific value of $\alpha$.

\begin{figure*}[thb]
\centering
\includegraphics[width=1\textwidth]{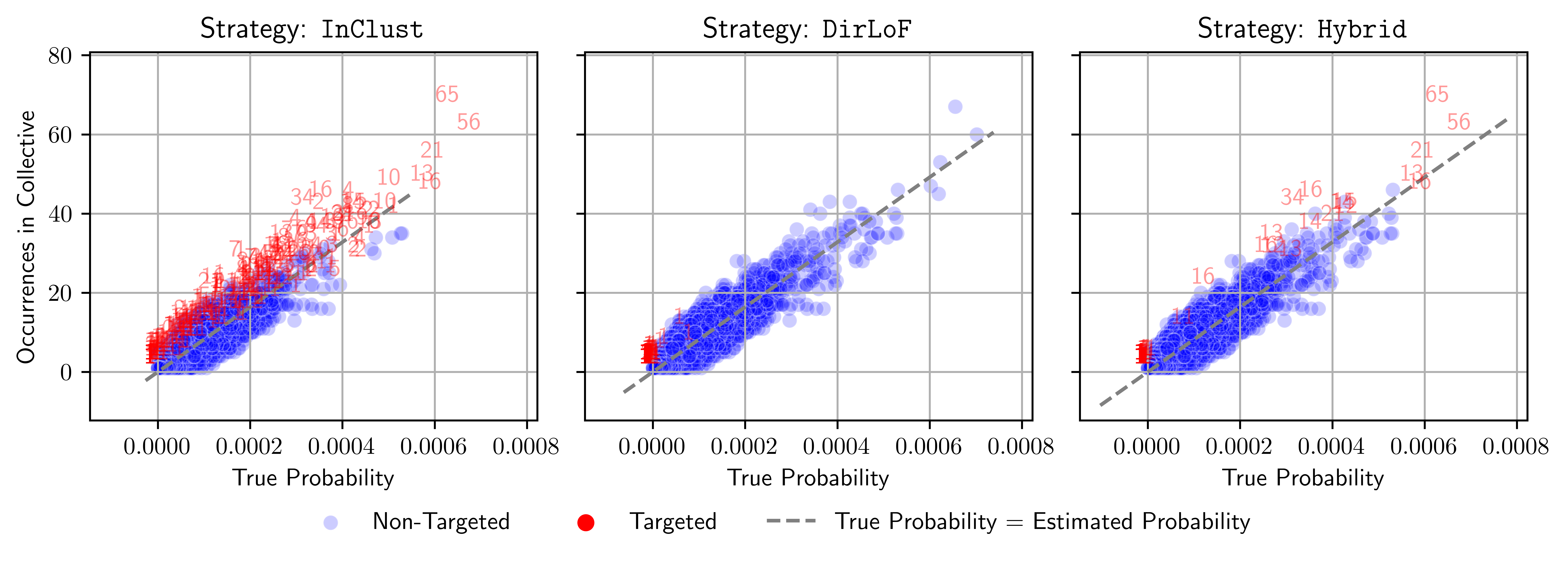}
\caption{
Songs in playlists owned by a collective composed of 0.07\% of the training data. Blue dots are songs that are not targeted. Red integers indicate a used anchor song and the number of times it is targeted.
}
\label{fig:targeted_anchors}
\end{figure*}

\subsection{Songs targeted by different strategies}

Figure~\ref{fig:targeted_anchors} illustrates the anchor song selection for three distinct strategies. The \texttt{InClust} strategy identifies anchors based on their prevalence within the collective, targeting songs frequently listened to by its members. Highlighted in red in the left panel of Figure~\ref{fig:targeted_anchors}, this method repeatedly employs indirect anchors, relying solely on internal playlist statistics without needing broader song frequency insights.
Conversely, the \texttt{DirLoF} strategy targets a specific cluster of direct anchors (resulting in the red-colored cluster of anchors in the bottom left corner of the middle panel in Figure~\ref{fig:targeted_anchors}) targeting each only once. This method requires external data to ensure the disproportionately represented anchors in the collective match those less prevalent in the training data, as shown on the x-axis. The \texttt{Hybrid} strategy, illustrated in the right panel of Figure~\ref{fig:targeted_anchors}, combines these approaches, targeting both frequently occurring songs within the collective and low-frequency anchors.
The parameter $\lambda$ governs the fraction of the collective targeting indirect anchors (left panel) versus targeting direct anchors (middle panel).
Larger $\lambda$ values result in a larger share of direct anchors being targeted, and smaller values result in keeping more of the frequently targeted indirect anchors, i.e., the large red integers visualized toward the top of the left panel.
Note that all strategies target several anchors, ensuring all playlists can be used effectively.

\subsection{Robustness of collective strategies}
\label{app:robustness}

Table~\ref{rebuttal:hp_robustness} shows that the \texttt{DirLoF} strategy is robust against hyperparameter (\texttt{hp}) changes.
It consistently outperforms the \texttt{Random} strategy across all configurations.
Finally, Table~\ref{rebuttal:nr_of_epochs} illustrates the effectiveness of \texttt{DirLoF} and \texttt{Random} across different numbers of training epochs for the recommender.
Notice that the effectiveness of the \texttt{DirLoF} strategy decreases if model training is stopped early.

\begin{table}[t!]
    \centering
    \caption{Mean Amplification (Std Dev) under different \texttt{hp} configurations ($\alpha = 0.001$).
    \texttt{hp$^*$} denotes the optimal set of \texttt{hp} reported by~\citet{Bendada2023DeezerPlaylistContinuationTransformers} with 8 attention heads (\texttt{n\_heads}), a learning rate (\texttt{lr}) of 1.0, a dropout rate (\texttt{drop\_p}) of 0.13, a weight decay (\texttt{wd}) of 1.53e-05, and 18 epochs.
    We experiment with five alternative \texttt{hp} configurations that are equivalent to \texttt{hp$^*$} except for the following changes:
    \texttt{hp1} sets \texttt{n\_heads}=4,
    \texttt{hp2} sets \texttt{lr}=0.5,
    \texttt{hp3} sets \texttt{drop\_p}=0.2,
    \texttt{hp4} sets \texttt{wd}=5e-05,
    and \texttt{hp5} makes all four of these changes.
    The highest amplification values among \texttt{hp} configurations are highlighted in bold.
    }
    \label{rebuttal:hp_robustness}
    \begin{tabular}{lllllll}
        \toprule
        & \multicolumn{6}{c}{Hyperparameter configuration} \\
        \cmidrule{2-7}
        Strategy & \texttt{hp$^*$} & \texttt{hp1} & \texttt{hp2} & \texttt{hp3} & \texttt{hp4} & \texttt{hp5} \\
        \midrule
        \texttt{Random} & 3.36 (4.45) & 1.12 (2.15) & 0.23 (0.52) & 2.63 (5.67) & 2.69 (2.96) & 2.69 (3.41) \\
        \texttt{DirLoF} & \textbf{16.38 (7.37)} & \textbf{18.47 (9.03)} & \textbf{21.77 (1.67)} & \textbf{16.13 (10.07)} & \textbf{26.32 (4.38)} & \textbf{27.92 (5.53)} \\
        \bottomrule
    \end{tabular}
\end{table}

\begin{table}[t!]
    \centering
    \caption{Mean Amplification (Std Dev) for relative to trained epochs ($\alpha = 0.001$). The best-performing strategies are highlighted in bold.}
    \label{rebuttal:nr_of_epochs}
    \begin{tabular}{lll}
        \toprule
        & \multicolumn{2}{c}{Strategy} \\
        \cmidrule{2-3}
        \# of epochs & \texttt{DirLoF} & \texttt{Random} \\
        \midrule
        1 & 0.02 (0.04) & \textbf{4.10 (9.17)} \\
        2 & \textbf{7.97 (5.86)} & 0.08 (0.17) \\
        4 & \textbf{9.88 (16.10)} & 2.17 (1.81) \\
        8 & \textbf{19.45 (12.82)} & 2.51 (4.82) \\
        12 & \textbf{25.79 (10.30)} & 4.23 (3.13) \\
        16 & \textbf{14.95 (6.00)} & 3.07 (4.85) \\
        18 & \textbf{16.38 (7.37)} & 3.36 (4.45) \\
        \bottomrule
    \end{tabular}
\end{table}

\subsection{Internalities and externalities of algorithmic collective action}
\label{app:In_and_Externalities_continuation}

Here we provide more details on the results presented in Section~\ref{ssec:Internalities_and_externalities}.

\paragraph{Metrics for model accuracy.}
\label{app:EffectonLearner}

To assess the quality of recommendations we follow~\citet{SpotifyRecSysChallenge2018} and~\citet{Bendada2023DeezerPlaylistContinuationTransformers} in using the following three popular performance metrics:
\textbf{R-precision} measures the fraction of recommended items present in the masked ground truth, augmented by artist matches. The Normalized Discounted Cumulative Gain (\textbf{NDCG}) measures the ranking quality by rewarding relevant tracks placed higher in the recommendation list. The number of clicks (\textbf{\#C}) quantifies how many batches of ten song recommendations (starting with top candidates) are needed to find one relevant track. The Deezer model trained on the unmanipulated data achieves comparable results to the winning solutions of the RecSys 2018 APC challenge along these metrics~\citep{SpotifyRecSysChallenge2018REPORT}.

\paragraph{Impact on other artists.}
\label{app:EffectonArtists}

Unlike the partially aligned interests of the firm and the collective, the dynamics among artists differ, since boosting recommendations for one artist inevitably reduces the visibility of others. We are interested in understanding who is affected by our strategy.
For the purpose of this analysis, we hold the total number of recommendations at inference time constant, making it a zero-sum game.

To complement Figure~\ref{fig:impact_on_other_artists}, in Table~\ref{tab:summary_statistics_impact_on_songs}, we show the effect of collective action (\texttt{hybrid} strategy with $\alpha=1\%$) on other songs.
In line with the result presented in Section~\ref{ssec:Internalities_and_externalities} (showing bins of songs with similar frequencies for just one fold), we do not find any evidence that any other songs experience a systematic change in exposure due to collective action, not even the targeted (in)direct anchor songs.

\begin{figure}[t!]
    \centering
    \includegraphics[width=0.4\textwidth]{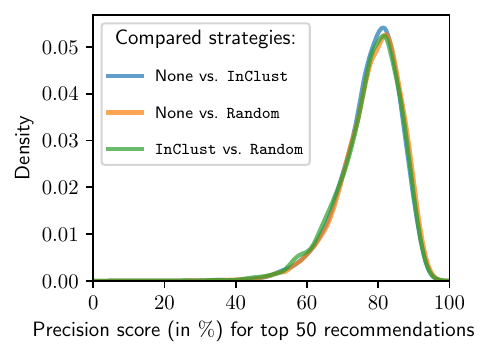}
    \caption{Variation in top 50 song predictions for attacked contexts: Predictions are stable under different collective action strategies.}
    \label{fig:ablation_prediction_stability}
\end{figure}

\begin{table}[t!]
\centering
\caption{Mean (± 95\% CI) recommendations without collective action ($R_0$), total gained recommendations ($\Delta R$), and gained recommendations in \% of $R_0$ (considering songs that are recommended at least once without collective action) for direct anchors, indirect anchors and others over five folds.}
\label{tab:summary_statistics_impact_on_songs}
\begin{tabular}{lccc}
\toprule
Metric & Direct anchors & Indirect anchors & Other songs \\
\midrule
$R_0$ & 0.09 ± 0.02 & 360.94 ± 11.29 & 0.29 ± 0.00 \\
$\Delta R$ & -0.00 ± 0.01 & 2.68 ± 3.74 & -0.00 ± 0.00 \\
$\Delta R$ in \% of $R_0$ & -0.00 ± 0.00 & 0.02 ± 0.01 & -0.00 ± 0.00 \\
Song counts (per fold) & 2784 & 157 & 2246756 \\
\bottomrule
\end{tabular}
\end{table}

\paragraph{User experience of collective participants}
\label{app:EffectonCollectiveParticipant}

Collective participants are platform users who continue consuming content on the platform during and after performing strategic actions. To understand the price of collective action, we investigate the downstream effect on their own user experience. After all, users are likely to engage in collective action only if it does not result in significant detriment to their own content consumption experience.
While a perfect measure of user satisfaction is outside the scope of this work, we utilize participant's known preferences as a basis for the experienced quality of recommendations.
More precisely, for any collective participant that manipulated their playlist with $h(p) = (p_{i^*}^-, s^*, p_{i^*}^+)$, we use the targeted context $(p_{i^*}^-)$ as a seed for the recommender and evaluate the output recommendations using the user-generated continuation of the playlist $(p_{i^*}^+)$ as the ground truth.

We revisit the collective strategy outlined in the ablation study (Section~\ref{ssec:AblationStudyIndirectContextEmbeddingAttacks}) to scrutinize the internalities of collective action.
We find that the recommendations for collective participants are stable and robust under authentic strategic playlist manipulations.
Figure~\ref{fig:ablation_prediction_stability} illustrates the precision score---measuring the similarity of recommendations for different strategies---across attacked contexts. It is around $80\%$ on average, with little variation across strategies. This is likely attributed to inherent instabilities in the top $K$ recommendations rather than the specifics of the strategies~\citep{Oh2022RankListSensitivity}.

Furthermore, by participating in collective action, users do not affect the variety of songs they get recommended.
This stability of model predictions despite the coordinated collective action is shown in Table~\ref{tab:ablation_prediction_robustness}: Inserting $s^*$ between $p_{i^*}^-$ and $p_{i^*}^+$ does not significantly distort the recommenders ability to predict $p_{i^*}^+$ from $p_{i^*}^-$.
Interestingly, random placement of songs reduces the performance slightly more (especially for \#C), as in this case, $s^*$ can be inserted within the context $p_{i^*}^-$ by pure chance.

\begin{table}[t!]
    \centering
    \caption{Average model performance: Predictions are robust under different collective action strategies.}
    \label{tab:ablation_prediction_robustness}
    \begin{tabular}{lccc}
    \toprule
                             Strategies &        NDCG & R-precision &       $\#$C \\
    \midrule
                            None & 0.35 ± 0.0 & 0.28 ± 0.0 & 1.43 ± 0.1 \\
    \texttt{Random} & 0.34 ± 0.0 & 0.28 ± 0.0 & 1.50 ± 0.1 \\
    \texttt{Inclust} & 0.35 ± 0.0 & 0.28 ± 0.0 & 1.42 ± 0.1\\
    \bottomrule
    \end{tabular}
\end{table}

\end{document}

%% file: macros.tex
\usepackage{xcolor}

\definecolor{DarkGreen}{rgb}{0.2,0.63,0.17}
\definecolor{DarkRed}{rgb}{0.89,0.1,0.11}
\definecolor{DarkBlue}{rgb}{0.12,0.47,0.71}
\definecolor{Gray}{rgb}{0.2,0.2,0.2}

\definecolor{DarkGreen}{RGB}{98,179,149}
\definecolor{DarkRed}{RGB}{206,86,69}
\definecolor{SABcolor}{RGB}{214,232,213}

\newcommand{\cP}{\mathcal{P}}
\newcommand{\cS}{\mathcal{S}}
\newcommand{\cD}{\mathcal{D}}

\DeclareMathOperator*{\argmax}{arg\,max}

\newcommand{\eat}[1]{}

\newcommand{\codeurl}{\url{https://github.com/joebaumann/recsys-collectiveaction}}